\documentstyle[11pt,epsf]{article}
\def\be{\begin{equation}}
\def\ee{\end{equation}}
\def\bea{\begin{eqnarray}}
\def\eea{\end{eqnarray}}

\title{ The Collective Excitation Spectra of $\sigma, \omega$ and  
$\pi$ Mesons in Nuclear Matter \footnote {The project supported
by the National Natural Science Foundation of China and Guangdong Province.}}
\author{Liu Liang-gang$^{1,2}$ \thanks{Email:stdp05@zsu.edu.cn} \\
$^1$Department of Physics, Zhongshan University, Guangzhou 510275, China\\
$^2$CCAST (World Laboratory), P O Box 8730, Beijing 100080, China }
\begin{document}
\maketitle
\date{}

\begin{abstract}

The recent progress on the study of the collective excitation in relativistic
nuclear matter is reviewed. The collective excitation modes are derived by 
meson propagators in nuclear matter. The mesons we study are $\sigma, \omega
, \gamma$ and $\pi$ mesons. For pion, we derived not only the relativistic
particle - hole, delta - hole excitations but also antiparticle excitations
, such as particle - antiparticle, antidelta - particle, delta - antiparticle
excitations. By calculating the dispersion relation and the spin - isospin
dependent response function, the effects of all these excitation are studied.
\\
{\vskip 1cm}
meson propagators, collective excitation spectra

\end{abstract}

\newpage
\section{Introduction}

Recently, the study if the hadron properties changing in nuclear matter
environment, such as the effective mass and so on, absorbs much attention
\cite{qm97}. It is reported that the experiment has shown a significant
amount of strength of the dilepton production spectrum below the free 
$\rho$ meson production threshold\cite{ceres,helios}, and this phenomena
can be explained by a downshift of the $\rho$ meson mass in nuclear matter
\cite{chanfray96}. There have been already a lot of theoretical studies
on the effective mass of $\rho$ meson and the light mesons $\omega, \phi$
in nuclear matter\cite{hatsuda,nakano,asakawa,herrmann,friman}. Not only
the vector mesons, the effective mass of $\sigma$ meson is studied also 
extensively (see, e. g., ref.\cite{saito}).

On the other hand, the effective mass of meson is related to the collective
excitation or dispersion relation of meson in nuclear matter. We have made 
a series of study on this subject for $\sigma, \omega$ meson
\cite{liu95a,liu95b} and pion\cite{liu97}. The collective excitation 
spectra of nuclear matter can be obtained by the trajectory of the pole
of meson propagators. In vacuum, the vacuum fluctuation caused by meson 
propagation will modify the dispersion relation of a free physical meson. 
In nuclear matter,  however, due to the
medium effects, the appearance of the dispersion relation  or the collective 
excitation spectra
will be very different from that in the vacuum, that is, in addition to the 
meson branch there will be acoustic sound excitation spectra in 
the space-like 
region. The meson branch can be regarded as the usual meson propagation in 
nuclear matter, it can also be interpreted as a mode of collective excitation 
which have the same quantum number as the meson. The acoustic sound spectra, 
disappears when nuclear density is zero, can only be viewed as the mode of 
collective excitation of nuclear matter\cite{chin}.

In this paper, we will review some of the progress in the studies of the 
collective excitation in nuclear matter. In the next section, we will
demonstrate the derivation of the relativistic particle - hole ({\it ph})
excitation in nuclear matter, then give the dimesonic function and collective
excitation spectra of $\sigma, \omega$ and electromagnetic field. In the 
3rd sect., we will study the spin - isospin dependent collective excitation
modes by pion propagator. The relativistic delta - hole ($\Delta h$)
excitation, antiparticle excitations and spin - isospin dependent response
function will be studied in the subsections. Finally, we give the summary
and conclusions in the last section.

\section{The collective excitation spectra of $\sigma, \omega$ mesons in
nuclear matter}

A systematic study of the collective excitation for $\omega$ meson 
in the Walecka
model has been given in ref.\cite{chin}.  
But as we have pointed out\cite{liu95a,liu95b,liu97} that the 
polarization insertion calculated in refs.\cite{chin}
, and in other papers does not
represent a real {\it ph} excitation neither a particle -  
antiparticle ($N \bar{N}$) excitation, because the nucleon 
propagator they used is "Feynman" propagator $G_{F}$ and "Density dependent"
propagator $G_{D}$, instead of particle and hole propagators. 
The polarization insertion calculated in that way, includes a finite $N \bar{N}$
excitation term which violates Pauli exclusion principle. The reason is that
a divergent $N \bar{N}$ excitation term is neglected in their calculation. In
our approach, however, the $ph$ and $N \bar{N}$ excitations are split up clearly.
The later is divergent and it is usually neglected in the nonrelativistic 
approach (see, for example\cite{oset}), 
but it can be taken into account when a appropriate renormalization
is made. We will demonstrate this point in the following by taking the example
of pion.

\subsection{Relativistic {\it ph} excitation}
For pseudo - scalar 
{\it PS} $\pi NN$ coupling, the polarization insertion ({\it PI}) 
of pion propagator in {\it NM} $\Pi^{PS}
(\equiv \Pi^{PS}_{F} + \Pi^{PS}_{D}$) calculated by using $G_{F}$ and $G_{D}$
propagators:
\bea
G(p) = G_{F}(p) + G_{D}(p)
\eea
\begin{eqnarray}
G_{F}(p) = \frac{1}{\not\!p - {\tilde m_N} + i \epsilon},
\end{eqnarray}
\begin{eqnarray}
G_{D}(p) = \frac{i \pi}{E_{\bf p}} (\not\!p + {\tilde m_N} 
) \delta (p_{0} - E_{\bf p}) n_{\bf p},
\end{eqnarray}
is as follows\cite{liu89}:
\begin{eqnarray}
\Pi^{PS}_{F}(q) & = &-  \frac{g^{2}_{\pi NN}}{(2\pi)^{3}} \int 
\frac{d {\bf p}}
{E_{{\bf p}}E_{{\bf p} + {\bf q}}} [4 E_{{\bf p} + {\bf q}} + \\ \nonumber
&  &q^{2} 
(\frac{1}{E_{{\bf p}} + E_{{\bf p} + {\bf q}} - \omega - i\epsilon} +
\frac{1}{E_{{\bf p}} + E_{{\bf p} + {\bf q}} + \omega - i\epsilon})],
\end{eqnarray}

\begin{eqnarray}
\Pi^{PS}_{D}(q)& =& \frac{g^{2}_{\pi NN}}{(2\pi)^{3}} \int \frac{d {\bf p}}
{E_{{\bf p}}E_{{\bf p} + {\bf q}}} \{(1 - n_{{\bf p}})n_{{\bf p} + {\bf q}}
\\ \nonumber &  & \cdot
q^{2}(\frac{1}{E_{{\bf p}} - E_{{\bf p} + {\bf q}} - \omega - i\epsilon} +
\frac{1}{E_{{\bf p}} - E_{{\bf p} + {\bf q}} + \omega - i\epsilon})  \\
\nonumber
&  & + n_{{\bf p}} [4 E_{{\bf p} + {\bf q}} + q^{2}(\frac{1}
{E_{{\bf p}} + E_{{\bf p} + {\bf q}} - \omega - i\epsilon} +
\frac{1}{E_{{\bf p}} + E_{{\bf p} + {\bf q}} + \omega - i\epsilon}) ]\},
\end{eqnarray}

here $p = (p_{0}, {\bf p}), 
q = (\omega, \bf q)$, $E_{{\bf p}} = \sqrt{
{\bf p}^{2} + \tilde{m}_N^{2}}$, $\tilde{m}_N$ is nucleon effective mass 
in nuclear matter. 
$n_{{\bf p}}$ is the nucleon distribution function at zero temperature. 
The subscripts $F$ and $D$ in above eqs. denote Feynman part and explicitly 
density dependent part, respectively. 
It is obvious $\Pi^{PS}_{F}$ is divergent, it comes from the contribution of
two $G_{F}$ in the integrand of the nucleon loop integration. People usually 
omit this term and keep $\Pi_{D}$ only (see ref.\cite{herbert}
, where $\pi NN$ coupling
is pseudo - vector ({\it PV})), and tread it as {\it ph} excitation. Later we
will see $\Pi^{PS}_{F}$ can not be treated as
$N \bar N$ excitation, neither $\Pi^{PS}_{D}$ is due to $ph$ excitation.

On the other hand, $G(p)$ can also be expressed as follows\cite{bentz}:
\begin{eqnarray}
G(p) = S_{p}(p) + S_{h}(p) + S_{\bar{p}}(p)
\end{eqnarray}
\begin{eqnarray}
S_{p}(p) = \frac{\tilde{m}_{N}}{E_{{\bf p}}} \frac{(1 - n_{{\bf p}}) \Lambda_{+}(
{\bf p})}{p_{0} - E_{{\bf p}} + i \epsilon},
\end{eqnarray}
\begin{eqnarray}
S_{h}(p) =  \frac{\tilde{m}_{N}}{E_{{\bf p}}} \frac{ n_{{\bf p}} \Lambda_{+}(
{\bf p})}{p_{0} - E_{{\bf p}} - i \epsilon},
\end{eqnarray}
\begin{eqnarray}
S_{\tilde{p}}(p) = - \frac{\tilde{m}_{N}}{E_{{\bf p}}} \frac{ \Lambda_{-}(-
{\bf p})}{p_{0} + E_{{\bf p}} - i \epsilon},
\end{eqnarray}
here $S_{p}$ is antinucleon propagator, 
$\Lambda_{+}({\bf p})$, $\Lambda_{-}({\bf p})$ is the particle and 
antiparticle projection operator, respectively. When these propagators are used to 
calculate {\it PI}, the nonvanishing parts will be due to  $ph$ and $N \bar N$
excitations, they are $\Pi^{PS}_{ph}, \Pi^{PS}_{N \bar N}$:

\begin{eqnarray}
\Pi^{PS} (q) = \Pi^{PS}_{ph} (q) +  \Pi^{PS}_{N \bar{N}} (q),
\end{eqnarray}
\begin{eqnarray}
 \Pi^{PS}_{ph} (q) &=& \frac{g^{2}_{\pi NN}}{(2\pi)^{3}}
\int \frac{d {\bf p}}{E_{{\bf p}}E_{{\bf p} + {\bf q}}} 
(1 - n_{{\bf p}})n_{{\bf p} + {\bf q}}[2(E_{{\bf p}} - E_{{\bf p} + {\bf q}})
\\ \nonumber
&  &+ q^{2}(\frac{1}{E_{{\bf p}} - E_{{\bf p} + {\bf q}} - \omega - i\epsilon} +
\frac{1}{E_{{\bf p}} - E_{{\bf p} + {\bf q}} + \omega - i\epsilon})],
\end{eqnarray}
\begin{eqnarray}
 \Pi^{PS}_{N \bar{N}} (q) & = &- \frac{g^{2}_{\pi NN}}{(2\pi)^{3}}
\int \frac{d {\bf p}}{E_{{\bf p}}E_{{\bf p} + {\bf q}}} 
(1 - n_{{\bf p}})[2(E_{{\bf p}} + E_{{\bf p} + {\bf q}})
\\ \nonumber
&  &+ q^{2}(\frac{1}{E_{{\bf p}} + E_{{\bf p} + {\bf q}} - \omega - i\epsilon} +
\frac{1}{E_{{\bf p}} + E_{{\bf p} + {\bf q}} + \omega - i\epsilon})],
\end{eqnarray}

The factor $(1 - n_{\bf p})n_{\bf p + \bf q}$ and  
$(1 - n_{\bf p})$ in above equations 
correctly indicated the excitations they belong.
It can be verified that the 
sum of $\Pi^{PS}_{ph}$ and $\Pi^{PS}_{N \bar N}$ is equal to sum of
$\Pi^{PS}_{F}$ and $\Pi^{PS}_{D}$. Since $\Pi^{PS}_{ph}$ is not equal to 
$\Pi^{PS}_{D}$, $\Pi^{PS}_{N \bar N}$ is not equal to $\Pi^{PS}_{F}$, so 
$\Pi^{PS}_{F}$ does not stand for $N \bar N$ excitation, neither 
$\Pi^{PS}_{D}$ stands for $ph$ excitation. That means, generally speaking, 
one can not get the correct expressions for 
$ph$ and $N \bar N$ excitation if one uses $G_{F}, G_{D}$ propagators to 
calculate {\it PI} in {\it NM}. Neglecting $\Pi_{F}$ and retaining $\Pi_{D}$ 
only in the calculation is not a matter of approximation or a truncation 
of a full
theory, explicitly speaking it is a wrong formula derived by a incorrect
approach. But that is not the case for $\Pi_{ph}$ and $\Pi_{N \bar N}$, 
because they are two independent excitation modes and the later is well 
known negligible in the nonrelativistic approximation\cite{oset}. 
So, we will no longer consider $\Pi_{F}$
and $\Pi_{D}$ again in the following formulae and numerical calculations.

\subsection{isoscalar longitudinal and transverse collective excitation}

The propagation of isoscalar $\omega$ meson in nuclear matter will cause
longitudinal and transverse collective excitation. It is well known that
the isoscalar $\sigma$ meson will mix with the longitudinal mode of 
$\omega$ meson\cite{chin}. To study these collective excitation modes, we 
start from the dimesonic function of nuclear matter\cite{liu95a}, it is
given as follows:
\begin{eqnarray}
\epsilon(q) = (\epsilon^{T}(q))^{2} \cdot \epsilon^{L}(q),
\end{eqnarray}
\begin{eqnarray}
\epsilon^{T}(q) = 1 - \Pi^{T}_{ph}(q) \Delta_{\omega}(q),
\end{eqnarray}

\begin{eqnarray}
\epsilon^{L}(q) = (1 - \Pi^{L}_{ph}(q) \Delta_{\omega}(q))(1 - \Pi^{S}_{ph}
(q) \Delta_{\sigma}(q)) + \frac{q^{2}}{{\bf q}^{2}}(\Pi^{0}_{ph}(q))^{2},
\end{eqnarray}
\begin{eqnarray}
\Delta_{\omega} (q) = \frac{-1}{q^{2} - m^{2}_{\omega}} , 
{\hskip 0.5cm} \Delta_{\sigma} (q) = \frac{1}{q^{2} - m^{2}_{\sigma}} , 
\end{eqnarray}
here $m_{\omega}, m_{\sigma}$ are the mass of $\omega$ and $\sigma$ mesons.
The last term in $\epsilon^{L}(q)$ is due to $\sigma - \omega$ mixing, and
\begin{eqnarray}
\Pi^{S}_{ph} (q)  = g^{2}_{\sigma}
[2C_{0}(q) + 
(q^{2} - 4 {\tilde m_N}^{2}) L_{0}(q)]
\end{eqnarray}
\begin{eqnarray}
\Pi^{L}_{ph} (q) = g^{2}_{\omega}[ 2 (\frac{C_{2}(q)}{{\bf q}^{2}} - 
C_{0}(q) ) + q^{2} ( \frac{L_{2}(q)}{{\bf q}^{2}} -  L_{0}(q) )],
\end{eqnarray}
\begin{eqnarray}
\Pi^{T}_{ph} (q) = - \frac{g^{2}_{\omega}}{2} [2 ( \frac{C_{2}(q)}
{{\bf q}^{2}} + C_{0}) + (q^{2} + 4 {\tilde m_N}^{2}) L_{0}(q) +
\frac{q^{2}}{{\bf q}^{2}} L_{2}(q)],
\end{eqnarray}
\begin{eqnarray}
\Pi^{0}_{ph}(q) = -2 g_{\sigma} g_{\omega} {\tilde m_N} L_{1} (q),
\end{eqnarray}
here $g_{\sigma}, g_{\omega}$ are the $\sigma NN$ and $\omega NN$
coupling constants, respectively,
\begin{eqnarray}
C_{n}(q) = 
\frac{1}{(2\pi)^{3}} \int \frac{d {\bf p}}
{ E_{\bf p} E_{\bf p - q}}
(1 - n_{\bf p}) n_{\bf p - q} (E_{\bf p} + E_{\bf p - q})^{n} 
(E_{\bf p} - E_{\bf p - q}),
\end{eqnarray}
\begin{eqnarray}
L_n (q) & = & 
\frac{1}{(2\pi)^{3}} \int \frac{d {\bf p}}
{ E_{\bf p} E_{\bf p - q}}
(1 - n_{\bf p}) n_{\bf p - q} (E_{\bf p} + E_{\bf p - q})^{n} 
\nonumber \\
&  &\cdot (\frac{1}{E_{\bf p} - E_{\bf p - q} - q_{0} - i \epsilon} +
\frac{1}{E_{\bf p} - E_{\bf p - q} + q_{0} - i \epsilon} ).
\end{eqnarray}

The analytic expression for the real and the imaginary part of 
$L_n(q)$ is given in ref. \cite{liu95c}.
The eigen condition for determining the collective excitation spectra is
$\epsilon(\omega, q) =0$.

There are three free parameters in the Walecka model, that is 
$g_{\sigma}, g_{\omega}, m_{\sigma}$. The effective mass of nucleon 
is given by the 
condition that the energy per nucleon should be the minimum to  the 
variation of nucleon effective mass. In ref.\cite{chin}, 
$m_{\sigma}$ = 550 MeV and
$g_{\sigma}, g_{\omega}$ are chosen to fit the binding energy of nuclear
matter at saturation density $k_{F}$ = 1.42 fm$^{-1}$.  The results are: in 
the mean field approximation (MFA) $\frac{\tilde m_N}{m_N}$ = 0.56, 
$g_{\sigma}$
 = 9.59, $g_{\omega}$ = 11.7. In the relativistic Hartree approximation (RHA)
$\frac{\tilde m_N}{m_N}$ = 0.72, $g_{\sigma}$ = 7.93, $g_{\omega}$ = 8.94. 
The masses of nucleon and $\omega$ meson are 938.5 MeV and 783 MeV, 
respectively. In Fig. 1. we show the results of collective spectra at
nuclear saturation density, calculated in the MFA. 
There are the meson branches
and acoustic sound branch indicated by letters "S" and "L", "T" which are
due to $\sigma$ meson, longitudinal and transverse $\omega$ meson. The 
longitudinal and transverse mode diverse at large energy and momentum, they
are almost overlay for low momentum. 

\begin{center}
\begin{tabular}{|c|c|} \hline
Fig. 1 &Fig. 2 \\ \hline
\end{tabular}
\end{center}

The acoustic sound spectrum forms a closed contour 
and most part of these acoustic sound 
spectra fall into the hatched region, namely, the corresponding modes are 
damped by $ph$ decays. In the RHA,  however, this acoustic sound mode
disappears as shown  in by Fig. 2,  the meson branch spectra only changes
slightly .

\section{Pion propagator in nuclear matter}

The pionic collective excitation and pion dispersion relation in nuclear
matter (or in finite nuclei) is very important to many fields in nuclear physics.
A typical example is the dilepton production in relativistic heavy - ion
collision in the energy region of $\rho$ meson production
\cite{gale,xia,chanfray93},
where the production rate is determined by the pion 
dispersion relation and its derivative. The vanishing derivatives of the 
dispersion relation curve may lead to a huge enhancement of the dilepton
production rate.
In ref. \cite{xia}, Xia et al calculated
the rate by using nonrelativistic pion dispersion relation which has very
good behavior for all momentum and energy. But it is not sure whether the 
nonrelativistic approximation is still applicable to very high energy - 
momentum transfer region. On the other hand, as far as pion dispersion 
relation is concerned, the imaginary part of the polarization insertion of 
pion propagator should not be ignored in the calculation.
In ref. \cite{liu97}, we made a first microscopic study of the {\it PI}
of the pion propagator and pion dispersion relation. We show in 
section 2.1 the relativistic {\it ph} excitation for {\it PS} $\pi NN$
coupling. In fact, pseudo-vector $\pi NN$ ({\it PV}) coupling is more 
frequently used instead of {\it PS} coupling. In this case, the {\it ph}
excitation {\it PI} can be written as follows:
\begin{eqnarray}
\Pi^{PV}_{ph} (q) = \frac{2f^{2}_{\pi NN}}{m^{2}_{\pi}}
[{\bf q}^{2}C_{0}(q) - C_{2}(q) + 2{\tilde m}_N^{2}
q^{2} L_{0}(q)],
\end{eqnarray}
here $f_{\pi NN}$ = 0.988 is {\it PV} $\pi NN$ coupling 
constant\cite{herbert}.

\subsection{Relativistic $\Delta h$ excitation}

The $\Delta h$ excitation is the most important excitation channel for pionic
excitation in nuclear matter. There have been many studies in the 
nonrelativistic formalism\cite{oset,helgesson}. The relativistic study begins
only recently\cite{liu97,herbert}, and in fact, only ref. \cite{liu97}
gives the correct results for relativistic $\Delta h$ excitation.

Similar to the nucleon propagator,
The delta baryon Feynman propagator $S^{\mu \nu}(p)$ can be expressed in 
terms of particle and antiparticle propagators, that is $S^{\mu \nu}_{\Delta}
(p)$ and $S^{\mu \nu}_{\bar \Delta}(p)$:
\begin{eqnarray}
S^{\mu \nu}(p) = S^{\mu \nu}_{\Delta}(p) + S^{\mu \nu}_{\bar \Delta}(p),
\end{eqnarray}
\begin{eqnarray}
S^{\mu \nu}_{\Delta}(p) = \frac{{\tilde m}_{\Delta}}{E_{\Delta}({\bf p})}
\cdot \frac{\Lambda^{\mu \nu}_{+}({\bf p})}{p_{0} - E_{\Delta}({\bf p}) 
+ i\epsilon},
\end{eqnarray}
\begin{eqnarray}
S^{\mu \nu}_{\bar \Delta}(p) = - \frac{{\tilde m}_{\Delta}}{E_{\Delta}({\bf p})}
\cdot \frac{\Lambda^{\mu \nu}_{-}( - {\bf p})}{p_{0} + E_{\Delta}({\bf p}) 
- i\epsilon},
\end{eqnarray}
where $p = (E_\Delta({\bf p}), {\bf p})$.
$E_{\Delta}({\bf p}) = \sqrt{{\bf p}^{2} + {\tilde m}^{2}_{\Delta}}$, 
${\tilde m}_{\Delta}$ is effective mass of $\Delta$-isobars in nuclear matter.
(Here we implicitly assume there is no real $\Delta$-isobars in the ground
state, otherwise the decay of a real $\Delta$-isobar will cause the instability of nuclear 
matter$^{19}$), for this reason we can omit the width of $\Delta$-isobars.)
$\Lambda^{\mu \nu}_{+}({\bf p})$, $\Lambda^{\mu \nu}_{-}({\bf p})$ is particle
and antiparticle projection operator, respectively\cite{liu97}.

The $\pi N \Delta$ interaction is given as follows\cite{herbert}:
\begin{eqnarray}
{\cal L}_{\pi N \Delta} = \frac{f_{\pi N \Delta}}{\sqrt{2} m_{\pi}}
{\bar \psi}^{\mu}_{\Delta}
{\vec T} \cdot \psi_{N} \partial_\mu{\vec \pi} + h.c.,
\end{eqnarray}
here $f_{\pi N \Delta}$ = 2$f_{\pi NN}$. 
Using $S_{h}$, $S^{\mu \nu}_{\Delta}$, ${\cal L} _{\pi N \Delta}$ 
and Feynman rules, $\Delta h$ polarization insertion can be given as follows:
\begin{eqnarray}
\Pi_{\Delta h}(q_{0}, {\bf q}) & = & - 
\frac{4f^{2}_{\pi N \Delta}}{9(2\pi)^3 {\tilde m}^{2}_{\Delta}
m^{2}_{\pi}} \int \frac{d {\bf p}} 
{E_{\Delta}({\bf p})E_{{\bf p - q}}} 
n_{\bf p - q} \{ 2(E_{\Delta}({\bf p}) - E_{\bf p - q}) 
\cr
& &[E_{\Delta}({\bf p})E_{\bf p - q} q^{2}_{0} 
+ {\tilde m}^{2}_{\Delta} q^{2} - ({\bf p} \cdot {\bf q})^{2}]
- 2 ({\tilde m}^{2}_{\Delta} - {\tilde m}^{2}_{N} - {\bf q}^{2}) \cdot
\cr
& &E_{\Delta}({\bf p}) q^{2}_{0} 
+[({\tilde m}_{\Delta} + {\tilde m}_{N})^{2}- q^{2}] 
[\frac{1}{2} (
(E_{\Delta}({\bf p}) + E_{\bf p - q} )^{2} + q^{2}_{0})
\cr
& &(E_{\Delta}({\bf p}) - E_{\bf p - q} )  
+({\tilde m}^{2}_{N} + {\bf q}^{2} -
{\tilde m}^{2}_{\Delta})(E_{\Delta}({\bf p}) + E_{\bf p - q} ) ]
\\ \nonumber
&  & + [q^{2} - ({\tilde m}_{\Delta} + {\tilde m}_{N})^{2}]
[{\tilde m}^{2}_{\Delta} q^{2} - \frac{1}{4}
({\tilde m}^{2}_{\Delta} - {\tilde m}^{2}_{N} +  q^{2})^{2}] \cdot
\cr
& &(\frac{1}{E_{\Delta}({\bf p}) - E_{\bf p - q} 
- q_{0} - i\epsilon} +
\frac{1}{E_{\Delta}({\bf p}) - E_{\bf p - q} + q_{0} - i\epsilon}) \}.
\end{eqnarray}

Shown in Fig. 3 is the real part of the {\it PI} for normal nuclear density

\begin{center}
\begin{tabular}{|c|c|} \hline
Fig. 3 &Fig. 4 \\ \hline
\end{tabular}
\end{center}

$\rho_0$. The parameters are determined in  RHA (the same for Fig. 4) without
$\Delta$ isobars, so we can set $\tilde m_\Delta = m_\Delta$. It is clear that
the {\it ph} excitation vanishes in the limit ($\omega, q$) $\rightarrow$ 0
but not the $\Delta h$ excitation. The short range correlation (SRC) effect is 
included in the standard way\cite{liu97,helgesson}, the Landau - Migdal 
parameter $g\prime$ is 0.6 in this calculation\cite{herbert}. The {\it ph}
excitation is suppressed by SRC in this case.

In Fig. 4, we show the pion dispersion relation in the RHA and $g\prime$ =
0.6 at $\rho_0$. The dispersion relation is obtained by the poles of pion
propagator. The curves are much more different from that in the relativistic 
approach. The curve e and the lower part of curve g represent $\delta h$ 
and $\pi$ branch, respectively. The curve d and the upper part of curve g is 
due to SRC effect. The curve f is a noncollective $\Delta h$ excitation mode, 
because it is mainly due to $\Delta h$ excitation contribution.

\subsection{$N\bar N, \bar \Delta N, \Delta \bar N$ excitation}
The one of the advantage of our relativistic approach is that the antiparticle
excitation can be separated from the conventional {\it ph} or $\Delta h$
excitation. Therefore, their effect can be studied separately. The real part 
of the {\it PI} of the antiparticle excitation is divergent and the 
renormalization is needed. For {\it PV} $\pi NN$ coupling, the divergence
can not be renormalized by adding counter terms. The imaginary part of
the {\it PI} describes the width of the corresponding decay process, is 
finite. The imaginary part of the {\it PI} for pion propagator is shown
in Fig. 5. The parameters are the same as for Fig. 4. For fixed momentum 
$q$ = 2.5k$_F$ (k$_F$ = 1.42 fm$^{-1}$), the threshold for different 

\begin{center}
\begin{tabular}{|c|c|} \hline
Fig. 5 &Fig. 6 \\ \hline
\end{tabular}
\end{center}

excitation is different but it is the same for $\Delta \bar N$ and 
$\bar \Delta N$ excitation. For {\it ph} and $\Delta h$ excitations, the
imaginary part is nonvanishing only for a finite energy range, but that is
not the case for antiparticle excitations where the imaginary part will not
vanish for any energy larger than the threshold, and the corresponding
amplitude is much larger than {\it ph} or $\Delta h$ excitation.

The imaginary part  relates to the real part by the dispersion relation
\begin{eqnarray}
\mbox {Re} \Pi (q_0, {\bf q})  =  \frac{P}{\pi} \int_0^\infty  d\omega^2 
{\mbox {Im} \Pi (\omega^2, {\bf q}) \over \omega^2 - q_0^2},
\end{eqnarray}
 
For antiparticle excitations, the imaginary part does not vanish for high 
energy and increases much faster than the denominator, so the integration
or the real part  is divergent. Our renoralization scheme is to make a 
form factor correction to the $\pi NN$ and $\pi N \Delta$ vertices, 
($f_{\pi NN}, f_{\pi N \Delta}$) $\rightarrow F_\pi (q^2) 
(f_{\pi NN}, f_{\pi N \Delta}$),
\begin{eqnarray}
F_\pi(q^2) = ({{\Lambda^2_\pi - m^2_\pi} \over 
{\Lambda^2_\pi - q^2}}) \theta (q^2 \le m^2_\pi) +
e^{- {{q^2 - m^2_\pi} \over \Lambda^2}} \theta (q^2 > m^2_\pi),
\end{eqnarray}
where $\Lambda_\pi$ = 1200 MeV, $\Lambda$ is a free parameter. Shown 
in Fig. 6 is the imaginary part of the response function of the charge
exchange reaction ($^3\mbox{He}$, t) with T$_{^3\mbox{He}}$ = 2.0 GeV
\cite{dmitriev85,dmitriev93}. We assume the density of the target nucleus
is $\rho$ = 0.17 fm$^{-3}$ and $\tilde m_N = m_N, \tilde m_\Delta =
m_\Delta$. The Landau - Migdal parameter $g\prime$ = 0.6. We can see that
for antiparticle 
excitation and form factor correction enhance the response function 
especially at resonance energy at about 260 MeV. Increasing cut - off 
$\Lambda$, (softening of the form factor for high energy) will increase 
the antiparticle excitation effect. We can also see that all three curves 
start from $\omega \sim$ 250 MeV which is the threshold of the $\Delta h$ 
excitation for $T_{^3{\mbox He}}$ = 2.0 GeV.

\section{Summary and conclusions}
The {\it ph} and $\Delta h$, etc., are the elementary and important
excitation modes in nuclear physics which were studied very well in the
nonrelativistic case. We propose a new formalism and applied to the 
relativistic case successfully. The relativistic {\it ph}, $\Delta h$
excitations are derived and the nonrelativistic results can be reproduced.
The antiparticle excitations, such as $N\bar N, \Delta \bar N, \bar \Delta
N$ excitation are derived. To our knowledge, we are the first derived the
antiparticle excitations. In this paper, we show the application of our
formalism to meson propagator propagators only, but in fact, it can be
generally applied to any of the processes including the {\it PI}. On the 
other hand, we are going to extend the formalism to finite temperature.

We thank H. Q. Chiang, M. Nakano, W. Chen, X. Q. Luo for valuable discussions
and cooperation. We are very grateful to the National Natural Science
Foundation (NNSF) of China for its continuous support. The project is also
supported by the Natural Science Foundation of Guangdong Province.
                                                                  

\newpage
\section*{Figure Captions}

\begin{description}
\item[Fig. 1:] 
The collective excitation spectra of $\sigma, \omega$ mesons
in the MFA at normal nuclear density. The letters S, L, T stand for $\sigma$
meson, longitudinal and transverse $\omega$ meson. The q and $\omega$ are in 
unit of $m_\omega$. 
\end{description}

\begin{description}
\item[Fig. 2:] 
The same as Fig. 1 but in the RHA.
\end{description}

\begin{description}
\item[Fig. 3:] 
The real part of the {\it PI} - Re $\Pi(\omega = 0, q)$ 
(in unit of $m_\pi^2$) as a function of momentum $q$ (in unit of $m_\pi$)
for nuclear density $\rho_0$. The {\it ph} and $\Delta h$ contributions are
shown by dotted and dashed lines. "ph + $\Delta$h + SRC" is the total
result.
\end{description}

\begin{description}
\item[Fig. 4:] 
 The pion dispersion relation calculated in the RHA at normal
nuclear density including {\it ph}, $\Delta h$ and SRC effect with $g\prime$
=0.6.  
\end{description}

\begin{description}
\item[Fig. 5:] 
-Im$\Pi (\omega, q)$ (in unit of $m_\pi^2$) versus energy
$\omega$ (in unit of $m_\pi$ for fixed momentum $q$ = 2.5 k$_F$. The meaning 
of the letters, which are explained in the text, indicate the corresponding
excitations.
\end{description}

\begin{description}
\item[Fig. 6:] 
The imaginary part of the spin - isospin dependent response function 
 Im ${\mbox  R}(\omega, q \equiv |{\bf q}|)$ (in unit of 
$m_\pi^2$) as a function of energy transfer $\omega$ (in unit of GeV).
The dashed and dotted lines are the result with cut - off parameter 
$\Lambda$ = 1.4 GeV and 1.5 GeV, respectively. The solid line curve 
corresponds to the case without antiparticle excitation contribution 
$\Pi_{N\bar N} = \Pi_{\bar \Delta N} = \Pi_{\Delta \bar N}$ = 0, and 
no form factor correction, namely $F_\pi (q^2_\mu)$ = 1.
\end{description}

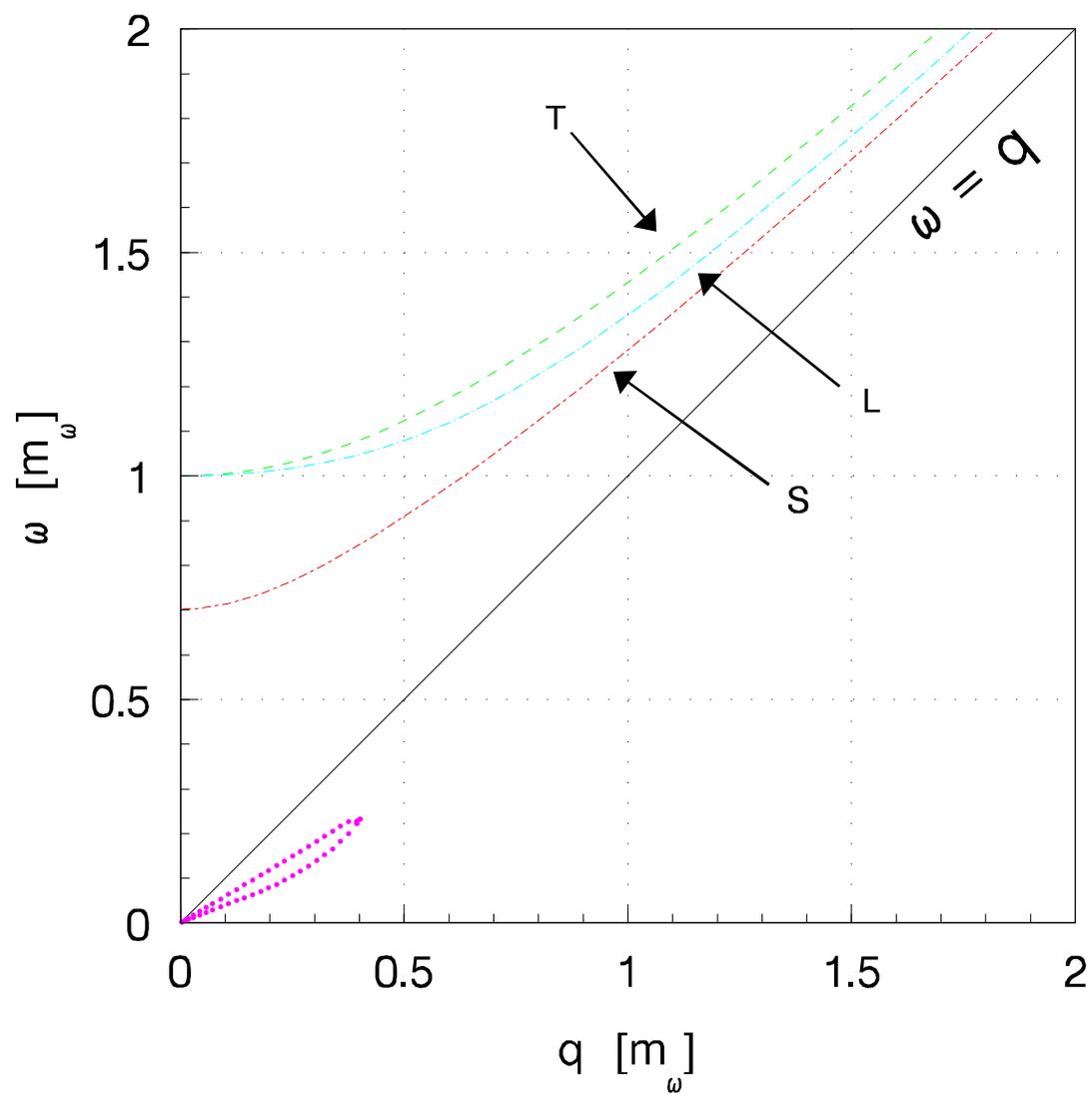
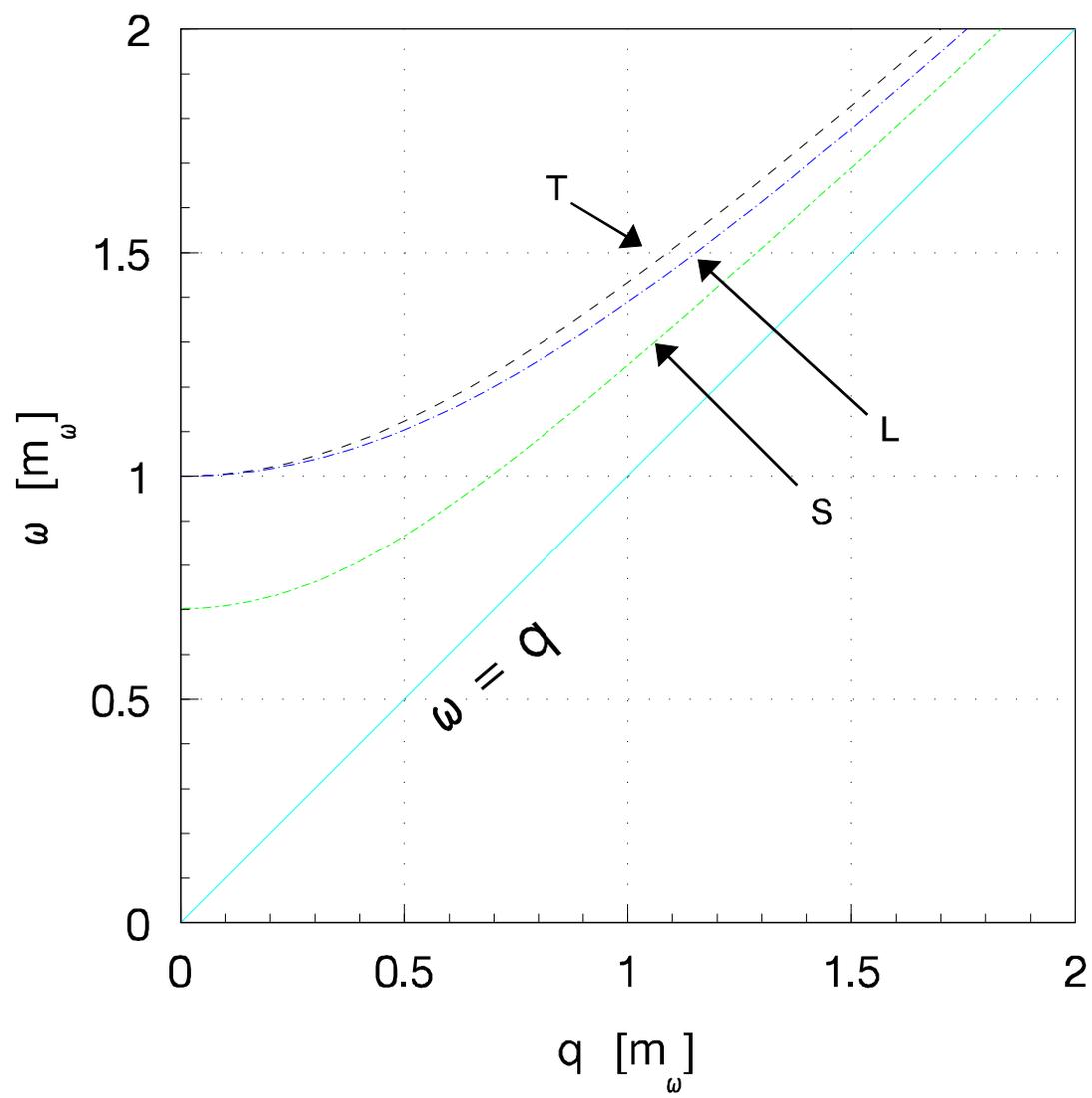
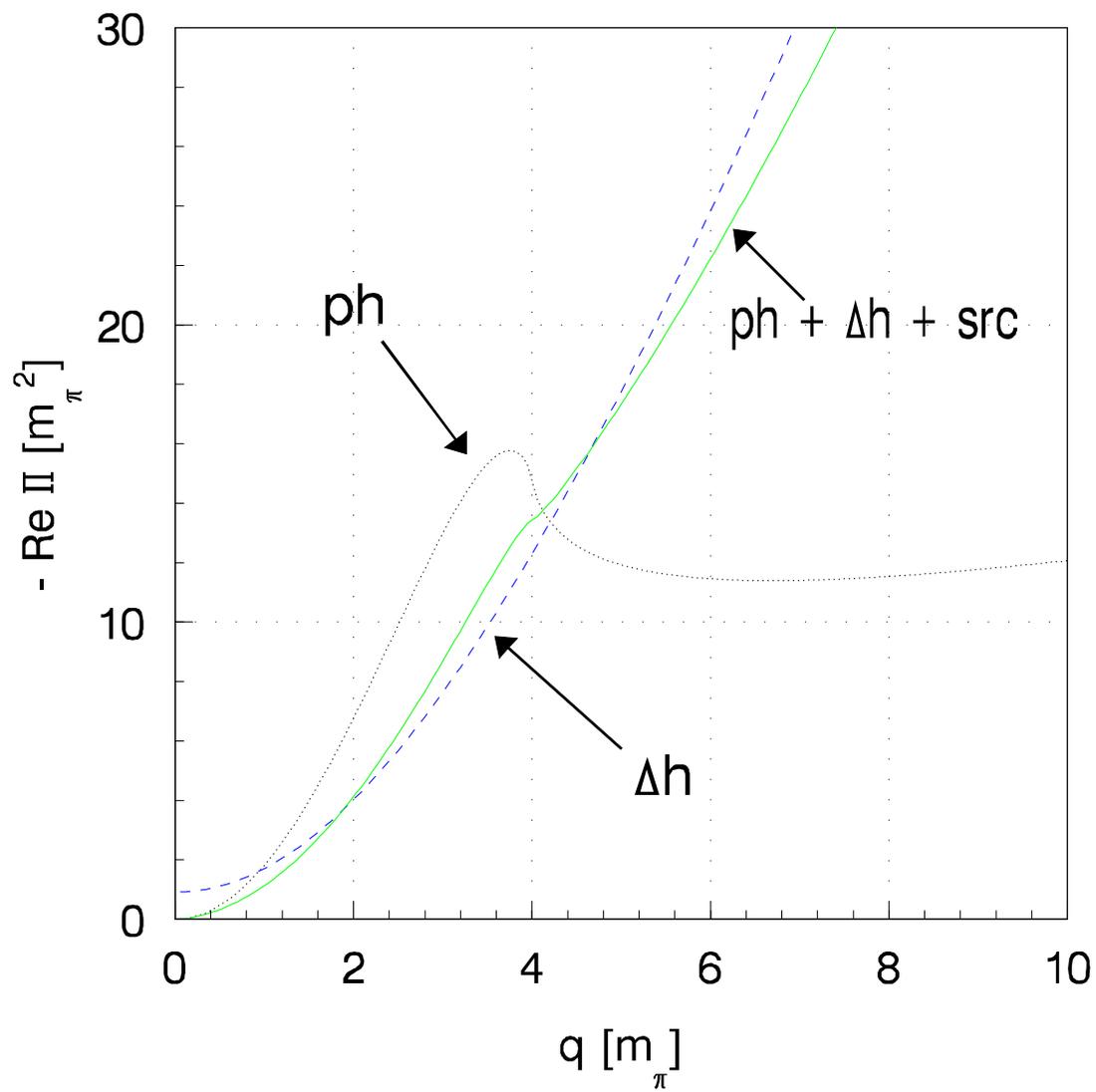
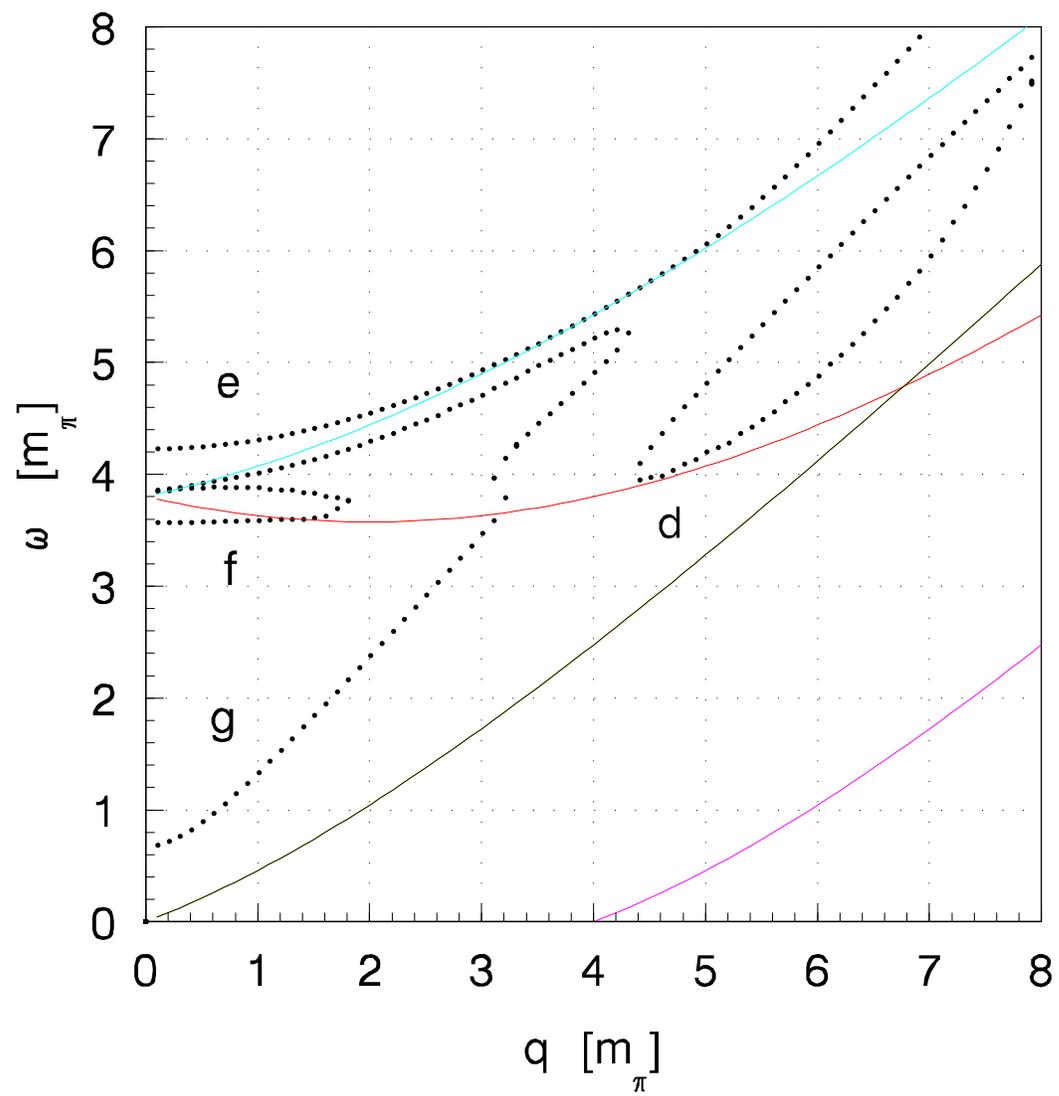
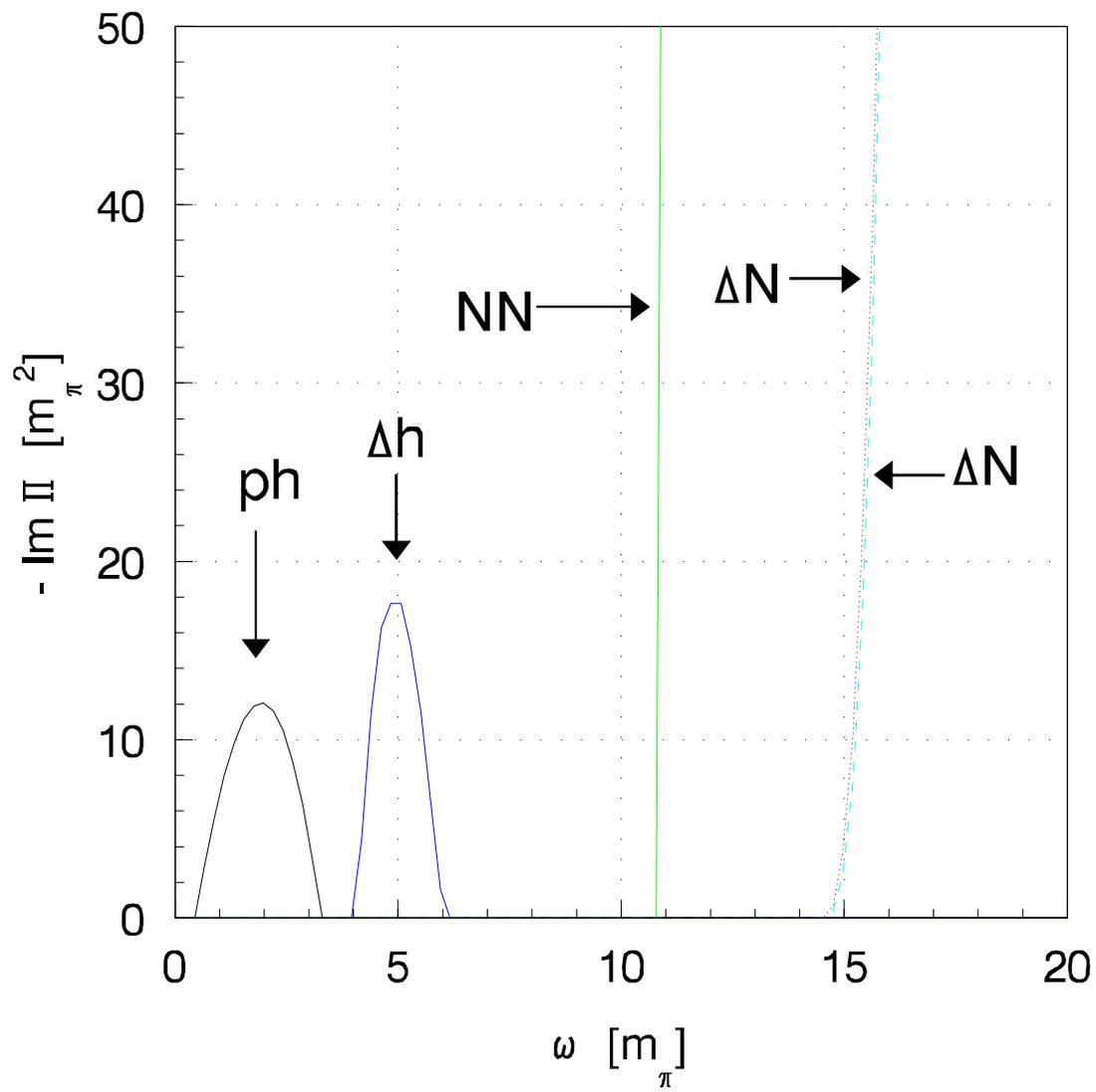
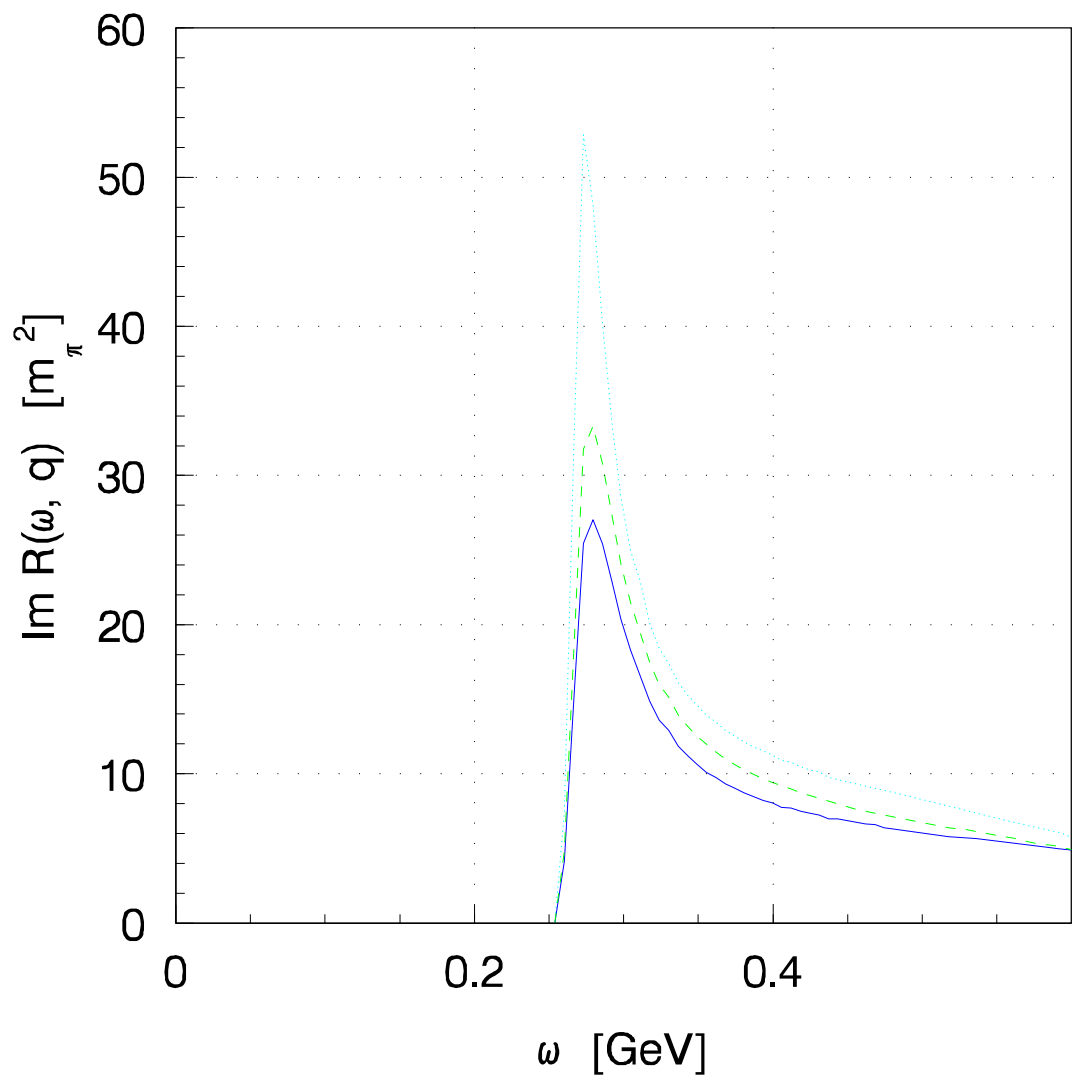
\end{document}